\documentclass[
  ,draft            % use draft while you are working on the paper
  ,numberedheadings % uncomment this option for numbered sections
]
  {aipproc}

\layoutstyle{6x9}

\newcommand{\smfrac}[2]{{\textstyle{#1\over#2}}}

\def\quarter{\smfrac{1}{4}}

\font\SYM=msbm10 % For blackboard bold

\def\Real{\hbox{\SYM R}}

\newcommand{\df}{{\mbox{\rm d}}}
\newcommand{\e}{{\rm e}}
\def\dfrac#1#2{{\displaystyle {#1 \over #2}}}
\newcommand{\vsa}{\vphantom{\dot{A}}} 
\newcommand{\ii}{{\rm i}}

\font\tenscr=rsfs10
\newcommand{\Scri}{{\mbox{\tenscr I}}}
\begin{document}

\title[Exact solutions of the Einstein equations]
 {Finding and using exact solutions of the Einstein equations}


\author{M.A.H. MacCallum}{
  address={School of Mathematical Sciences, Queen Mary, University of
  London, Mile End Road, LONDON E1 4NS, U.K.\\
  Email: m.a.h.maccallum@qmul.ac.uk}
}

\begin{abstract}
  The evolution of the methods used to find solutions of Einstein's
  field equations during the last 100 years is described. Early papers
  used assumptions on the coordinate forms of the metrics. Since the
  1950s more invariant methods have been deployed in most new papers.
  The uses to which the solutions found have been put are discussed,
  and it is shown that they have played an important role in the
  development of many aspects, both mathematical and physical,
  of general relativity.
\end{abstract}

\maketitle

%%%%%%%%%%%%%%%%%%%%%%%%%%%%%%%%%%%%%%%%%%%%
%% MAINMATTER
%%%%%%%%%%%%%%%%%%%%%%%%%%%%%%%%%%%%%%%%%%%%

\section{Introduction}

In an hour's talk such as this it is impossible to cover all that is
known on the subject, especially since in some respects detail is of
the essence in dealing with exact solutions. The three major recent
reviews of rather general character \cite{SteKraMac03,Kra97,Bic00}
have a total of over 1200 pages and that is before moving to more
specialized reviews such as \cite{Gri91,BelVer01}. (The short summary
of what is new in \cite{SteKraMac03} as compared with the first
edition, given below as an appendix, illustrates some ways in which
the field has changed in the last 25 years.) Thus I shall be
selective, dwelling at some length on particular solutions rather than
trying to cram in as many as possible in the time. For example, I give
particular attention to the Schwarzschild solution, the first solution
known that did not have constant curvature. I apologize to the authors
of the many excellent pieces of work I do not touch on.

Before learning the techniques for finding solutions of the Einstein
equations and discovering their properties, one should ask ``why this
is a worthwhile endeavour?''. Some colleagues do seem to regard it as
something of a backwater in the theory. The reason it is still
important stems from the nonlinearity of general relativity, one of
its essential features. To understand the meaning of the theory, there
are really three approaches. One can seek to prove global results,
such as are described in \cite{HawEll73} and were the subject of the
Isaac Newton Institute programme in progress at the time of writing
(see \url{http://www.newton.cam.ac.uk/webseminars/pg+ws/2005/gmr/}).
One can try to use approximation, either in the form of iterated
perturbation methods or numerical solutions. Finally, one can use
exact solutions: as Mason and Woodhouse said, ``they combine
tractability with nonlinearity, so they make it possible to explore
nonlinear phenomena while working with explicit solutions''
\cite{MasWoo96}, and as we shall see below, they have had considerable
impact on the theory.

We also need to ask: ``what is a solution?''. We could assume a form
for the metric, calculate its Einstein tensor, and so obtain a form
for the energy-momentum through Einstein's equations. The
pointlessness of such `solutions' was made clear by Synge
\cite{Syn60}. More generally, choosing a more complicated form of
energy-momentum with a simple form of metric usually reduces the
number of equations to be solved and makes the task of finding
solutions easier. The vacuum case, and cases with an equivalent set of
equations to actually solve, are more difficult.

Of course, exact solutions are very special cases. However, as Bicak
\cite{Bic00} noted, Feynman said ``The physicist is always interested
in the special case. He is talking about something, he is not talking
abstractly about anything. He wants to discuss the gravity law in
three dimensions: he never wants the arbitrary force case in $n$
dimensions. So a certain amount of reducing is necessary...''.  The
special cases known exactly can be very useful as examples in guiding
the global or approximative approaches.

Indeed without such uses finding exact solutions would be more like
stamp-collecting than science. It requires ingenuity and the objects
found may be beautiful, which is fine if you think relativity is a
branch of pure mathematics, but if you think it is physics, that is
not so good, despite Feynman's point. It is not the formulae for the
solutions that are really of interest: ``At present the main problem
concerning solutions, in our opinion, is not to construct more but
rather to understand more completely the known solutions with respect
to {\em their local geometry, symmetries, singularities, sources,
  extensions, completeness, topology and stability}'' (my emphasis).
This remark by Ehlers and Kundt in 1962 \cite{EhlKun62} is as true now
as when it was written, perhaps more so as a result of the large
number of further solutions found since then.  Part of my
purpose here is to illustrate the crucial role such understanding of
exact solutions has played in the development of our understanding of
general relativity itself\footnote{These aspects are covered at
  greater length in Bicak's review \cite{Bic00}, which I found very
  useful when selecting points to cover below.}.

The illustrations of uses of specific solutions given below naturally
focus on some of the best-known solutions. One might imagine that the
very large number of other solutions known have little use. For many
that may be true, at least so far, but many others have helped the
exploration of the physics.

The effort required to find solutions is not trivial. Compared with
Newtonian gravity, general relativity has one more independent
variable, 9 more dependent ones (taking the metric approach), and
equations of degree 8 in these variables rather than one, the result
of these changes being that the general form of the field equations
expands to 10 partial differential equations each with thousands of terms
(in terms of the metric and coordinates). This complexity is the
reason only some special solutions can be found, principally those
with some special symmetry or algebraic property.

One might nevertheless hope for a general solution. The closest to this which I
know of (the formalism of \cite{SciWayGil69}) is however largely
intractable. So I feel hopes for a useful general solution are slim
and instead one has to make the simplifying assumptions already
mentioned. Here the choices have developed over the last 90 years.

One disadvantage of those choices is that different ones may lead us
to the same solution. I repeat a jeu d'esprit I first gave in a
review some years ago \cite{Mac89}, by giving a list of the most
commonly-rediscovered solutions:
\begin{enumerate}
\item Flat space
\item The Schwarzschild solution
\item The Kasner solutions
\item Plane waves
\item Conformally flat perfect fluids
\item The Taub-NUT family of solutions
\item Static spherically symmetric perfect fluids
\item Cylindrically symmetric stationary electrovac solutions
\item Plane symmetric fluid and electrovac solutions
\item Spherically symmetric shearfree fluids
\end{enumerate}
The 7th and subsequent places in this list are actually rather open to
debate, for example the Harris Zund class are strong contenders. I will
say something later about how to recognize known solutions.

I should add that in this review I am going to stick firmly to the
number of dimensions that I know I exist in, i.e.\ 4, and avoid trespassing
on the 5- and higher-dimensional work described by others.  The
development of exact solutions in such theories seems to me to be for
the most part still at the stage of using only very simple metric
forms.

\section{The pre-existing solutions}

Since the title of this meeting is `A century of relativity', not the
90 of General Relativity, I am obliged to begin at the beginning,
meaning the first solution, the spacetime of special relativity,
Minkowski space:
\begin{equation}
\df s^2 =\df x^2+\df y^2+\df z^2 - \df t^2.
\end{equation}
It is flat, empty and is usually taken to have $\Real^4$ topology.

The role of this solution has been considerable, for example:
\begin{itemize}
\item It provides the prototype for asymptotic flatness (see Ehlers'
  contribution to this volume)
\item It is ideal for `cutting and pasting', which was a technique
  used to great effect while the concepts of causal structure were
  being developed\footnote{As a student I had the pleasure of watching
  part of this work as member of Sciama's research group in Cambridge.}
\item It was the setting for the first work on acceleration horizons
  (the Unruh effect)
\item In suitable coordinates, it is the Milne universe which is often
  used as the extreme Robertson-Walker model with $k= -1$. Apart from
  the use of this example in astrophysical predictions, this second
  choice of slicing of flat space, together with the three foliations
  of de Sitter space, helps to illustrate the fact that the 'open',
  'flat' or 'closed' nature of spacetimes can depend on the slicing.
\item It is a special case of many other models, whence its frequent
  rediscovery
\item It is the main background for quantum field theory (QFT), and plays a
  role even for QFT in curved spaces.
\end{itemize}

The other solutions one could regard as in some sense known before GR
was discovered were the other spaces of constant curvature, the de
Sitter and anti-de Sitter spaces:
\begin{equation}
 \df s^2=\frac{ \df x^2+  \df y^2 + \df z^2  - \df  t^2}
 {\left[ 1+\quarter K(x^2+y^2+z^2-t^2) \right]^2},%            \label{eq:8.35}
\end{equation}
where $K=\pm 1$.
These have played a role in, for example,
\begin{itemize}
\item Inflation (for de Sitter)
\item The AdS/CFT correspondence (anti de Sitter)
\item The ``no hair'' theorems (de Sitter)
\item The development of our understanding of particle horizons and
  event horizons.
\end{itemize}

Bicak \cite{Bic00} notes that stability against general non-linear
vacuum perturbations has been proved for these three solutions. This
is a `robustness' result: it raises the question of how much of what
we do is robust in this sense.

\section{Finding solutions: the first phase}

\subsection{Methods used}

Up to the 1950s the main method of finding new solutions started by
postulating a coordinate form of the metric. Authors generally
assumed one of:
\begin{itemize}
\item spherical symmetry,
\item cylindrical symmetry,
\item staticity or stationarity and axisymmetry, or
\item a plane wave form.
\end{itemize}
In the large bibliography amassed as background for writing
\cite{SteKraMac03}, only a part of which appears in the book itself, I
found that nearly all papers before 1950 belonged to one of these
groups, by far the largest group being the spherically symmetric
cases.  It should be noted that the metric forms were usually just
written down, whereas now we would derive them from
group-theoretic or other invariant assumptions.

As successful examples of this method, one can cite the work of
Schwarzschild, Droste, Levi-Civita, Kasner, Chazy, Curzon, Brinkmann,
Baldwin and Jeffrey, Lewis, van Stockum, Papapetrou, and G\"odel, refs.\
\cite{Sch16} --
\nocite{Dro16,Lev17,Wey17,Kas21,Cha24,Cur24,Bri25,BalJef26,Lew32,van37,Pap47,Pap53}\cite{God49},
and of course the Robertson-Walker metrics. Many of these solutions
have been important in later discussions, for example in establishing
the reality of gravitational waves or elucidating the nature of
directional singularities, but I will focus only on the two
best-known.

\subsection{The Schwarzschild solution}

The original form of the Schwarzschild solution \cite{Sch16} used a radial
coordinate $r$ which had its origin at the horizon: for clarity I
denote this $r_0$ below. Schwarzschild also
gave (contrary to some statements in the literature) the form most
often quoted now:
\begin{equation}
\label{Schw}
\df s^2=r^2(\df \vartheta ^2+\sin ^2\vartheta \,%
\df \varphi ^2)+A^{-1}\df r^2-A\df t^2,
\end{equation}
where $A=1-2m/r$.  Here I have renamed Schwarzschild's $R$ as $r$.
Schwarzschild regarded this as an auxiliary form because it did not
fulfil the condition $|\det (g)| = 1$ which Einstein had imposed in
the initial formulation of General Relativity, a condition of course
later discarded.

This solution initiated a discussion (see e.g.\ 
\cite{Eis82,Eis87}) on the meaning of the surface $r=2m$ and the absence of
a solution clearly analogous to a point mass in Newtonian theory.
There is of course no alternative point mass solution to consider due
to the uniqueness of (\ref{Schw}) as the spherically symmetric vacuum
solution (Birkhoff's theorem). 

There are still authors who argue that Schwarzschild's original
$r_0=0$ should be regarded as a singularity representing a point mass.
The horizon clearly is not a regular point since the area of
surrounding spheres has a limit $4\pi(2m)^2$ (which is part of the
reason $r=2m$ is now understood as a sphere, not a point). The
argument can be made at various levels of sophistication, the simplest
being that the metric components are singular (similar things could be
said of the axis of spherical polars, but nobody argues this is
singular!). To counter such arguments, it helps to note first that the
horizon $r=2m$ ($r_0=0$) is not in the coordinate patch for
(\ref{Schw}). (The waters here have been somewhat muddied by Hilbert's
arguments that these coordinates could be continued to the interior,
in which he overlooked that at the horizon the three coordinates
$(t,\, \theta,\, \varphi)$ do not parametrize a three-dimensional
manifold, as becomes immediately apparent on passing to the
Kruskal-Szekeres picture.) The whole Schwarzschild patch $r>2m$ is
isometric to a region of the Kruskal-Szekeres solution (region I in
the conformal diagram given as Fig. 1),
\begin{equation}
\df s^2  =  r^2(\df \vartheta ^2+\sin ^2\vartheta\,\df %
\varphi ^2)-32m^3r^{-1}\e ^{-r/2m}\df u\,\df v,
\end{equation}
where $r$ is defined implicitly by the equations giving $u$ and $v$ in
terms of the previous $t$ and $r$:
\begin{equation}
u  =  -( r/2m-1) ^{1/2}\e ^{r/4m}\e ^{-t/4m},
\quad v=(r/2m-1)^{1/2} \e ^{r/4m}\e ^{t/4m}.
\end{equation}
By inspection of this form of the solution we see that considering the bounding
surface $r=2m$ at a given time as a point amounts to topologically
identifying all the points on a sphere. One can produce an analogous
effect by cutting and pasting flat space (in $r_0$-like coordinates).
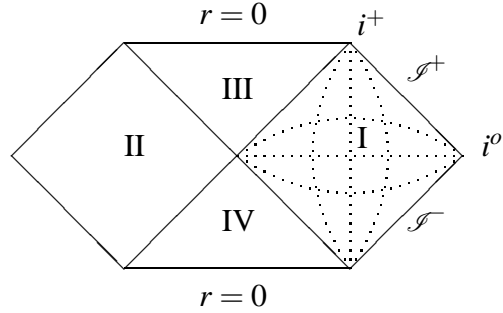
\begin{figure}
\setlength{\unitlength}{1 truecm}
\begin{picture}(8,5)
\put(2.5,4){\line(1,0){3}}
\put(2.5,1){\line(1,0){3}}
\put(2.5,4){\line(1,-1){3}}
\put(2.5,1){\line(1,1){3}}
\put(1,2.5){\line(1,-1){1.5}}
\put(1,2.5){\line(1,1){1.5}}
\put(5.5,4){\line(1,-1){1.5}}
\put(5.5,1){\line(1,1){1.5}}
\put(5.6,2.6){I}
\put(2.5,2.5){II}
\put(3.8,3.25){III}
\put(3.8,1.5){IV}
\put(7.25,2.5){$i^o$}
\put(5.6,4.1){$i^+$}
\put(6.25,3.5){$\Scri^+$}
\put(6.25,1.5){$\Scri^-$}
\put(3.5,4.25){$r=0$}
\put(3.5,0.5){$r=0$}
\qbezier[25](4,2.5)(5.5,3.5)(7,2.5)
\qbezier[25](4,2.5)(5.5,2.5)(7,2.5)
\qbezier[25](4,2.5)(5.5,1.5)(7,2.5)
\qbezier[25](5.5,1)(4.5,2.5)(5.5,4)
\qbezier[25](5.5,1)(5.5,2.5)(5.5,4)
\qbezier[25](5.5,1)(6.5,2.5)(5.5,4)
\end{picture}
\caption{Conformal diagram of the Kruskal-Szekeres form of the
  Schwarzschild solution. Each point shown represents a two-sphere
  parametrized by $\theta$ and $\varphi$. Coordinates $u$ and $v$ are
  constant along lines at $45^o$. The $45^o$ lines crossing in the
  centre of the figure are the horizon $r=2m$. The $45^o$ lines at the
  sides represent null infinity $\Scri$. Region I is isometric to the
  Schwarzschild region $r>2m$. Region II is a second exterior. Region
  III is the black hole interior and region IV a white hole
  interior. The dotted curves running to $i^o$ represent surfaces of
  constant $t$, and the ones through $i^+$ surfaces of constant $r$.
  Note that for all $t$, $r=2m$ is the single point at the centre,
  i.e.\ a single sphere.}
\end{figure}

One thing to note in these arguments is the ambivalence with which we
treat coordinates. Introductions to General Relativity always
emphasize the covariance of the equations, but practical examples often
implicitly communicate the importance of particular coordinates. The
problems this causes are seen at their worst when refereeing weak
papers on exact solutions, where authors often refuse to accept that
their solution is not new on the grounds its coordinate form is
different from the known ones.

The full understanding of the Schwarzschild horizon was an important
element in the general understanding of global properties of
spacetimes which developed in the 1960s. Conformal pictures drawn on
the same principles as Fig. 1 were also developed for the
Schwarzschild solution's generalizations given
by
$$A=1 -2m/r +e^2/r^2 -\smfrac{1}{3}\Lambda r^2,$$
the
Reissner-Nordstr\"om solution being given by $\Lambda =0 \neq e$ and
the K\"ottler solution by $\Lambda \neq 0 = e$. Incidentally, these
provide two illustrations of a `metatheorem' that all named results
have the wrong name: the K\"ottler solution is commonly called
Schwarzschild-anti-de Sitter, though as far as I know neither
Schwarzschild nor de Sitter gave this solution, and Weyl gave the
general class to which the Reissner-Nordstr\"om metric belongs before
Nordstr\"om's paper appeared. Other examples of this are the
``Tolman-Bondi'' solutions, due to Lema\^{\i}tre \cite{Lem33}, whose
paper Tolman \cite{Tol34} cited, and the ``Bertotti-Robinson''
solution, which Robinson himself noted was first found by Levi-Civita
\cite{Lev17}.

The Schwarzschild solution had a pivotal role in two other
developments. Approximations to it were used in predictions of the
classical tests of general relativity: those approximations were later
generalized to the PPN formalism which provides the basis of analysis
of solar system tests of relativity. It was also the setting in which
Hawking made the link between the laws of black hole mechanics,
quantum field theory and thermodynamics in his discovery of Hawking
radiation.

The thermodynamic identification of black hole surface area with
entropy is in general not fully understood in terms of microscopic
states, unlike entropy in normal statistical mechanics (see
\cite{Wal01}). However, for the `extreme' charged case\footnote{I am
  grateful to Ingemar Bengtsson and Gary Horowitz for correcting a
  misapprehension on my part about this work.} (which is
supersymmetric) or near-extreme charged black holes, a microscopic basis in
string theory has been given \cite{Hor98}. Yet again the Schwarzschild
solution's generalizations are playing a major role in advancing our
understanding.

To summarize, the Schwarzschild solution assisted the development of
our understanding of the following:
\begin{itemize}
\item there is no ``point mass''
  solution in relativity,  since there is no point centre;
\item The role of coordinates has to be properly understood;
\item global concepts such as\\
black holes,\\
event horizons,\\
apparent horizons,\\
trapped surfaces and the singularity theorems,\\
cosmic censorship and naked singularities;
\item PPN expansions; and
\item Hawking radiation, QFT in curved spaces and black hole entropy.
\end{itemize}

\subsection{The Robertson-Walker metric form: FLRW solutions}

These are the solutions with the metric
\begin{equation}
\label{eq:N12.1}
\df{s^2}=-\df{t^2} + a^2(t)[\df{r^2}+ \Sigma^2(r,k)
(\df{\vartheta^2}+\sin^2\vartheta\, \df{\varphi^2})].
\end{equation}
where
\begin{equation}
\hspace*{-1em}
\Sigma(r,k) = \sin r,\,r~\mbox{\rm or}~\sinh r, \mbox{\rm respectively, when}~k=1,\,0~\mbox{\rm or}~ -1.%\label{eq:N8.38}
\end{equation}

It would be impossible to overemphasize the importance of these
solutions in cosmology. They are fundamental to the inferences from
observation of the presence on the cosmological scale of dark matter
and ``dark energy'' (above the densities required by dynamics of
galaxy clusters). These matters were discussed by other speakers at
the meeting so I will not give details here.

There are many specific solutions due to Einstein himself, de Sitter,
Friedman, Eddington, Lema\^{\i}tre, and so on (see Ch.\ 14 of
\cite{SteKraMac03}). The pivotal role of the ones due to Friedman and
Lema\^{\i}tre led to those authors' names being coupled with the names
of Robertson and Walker in the short name FLRW. The role of Robertson
and Walker was that they independently elucidated the geometrical
basis of the metric form in the 1930s.

One may also note the wide use of the Lema\^{\i}tre-Tolman-Bondi models,
spherically symmetric inhomogeneous dust models, to describe collapse and
voids, primordial black holes, and other inhomogeneities in cosmology
(see \cite{Kra97}). For example, it has been shown that the number
distribution of galaxies can be modelled by a non-evolving population
in an LTB spacetime rather than an evolving population in FLRW.

\section{Finding more solutions: the second phase}

\subsection{Two major themes}

Two very big steps forward were made in the early 1950s, which
provided the framework for many of the new solutions found up to now.

The first of these was Taub's 1951 paper \cite{Tau51} which contained
the introduction of group-theoretic and differential geometric methods
used in an essential way. (There was also some interaction with
parallel work of G\"odel.) This paper also discovered the Taub portion
of the important Taub-NUT solution, and hinted at tetrad methods,
though these latter were not fully developed until the 1960s, by Ellis
\cite{Ell64,Ell67}, Estabrook and Wahlquist \cite{EstWah64}, Newman
and Penrose \cite{NewPen62} and others (a development which I helped
to codify \cite{Mac73} as far as the orthonormal tetrad version was concerned).

The second was Petrov's classification of the Weyl tensor
(\cite{Pet54}). This, together with the solutions of the Kundt class,
played an important part in the development of gravitational radiation
theory and formed a big step on the road to invariant classification
of solutions, discussed later.

One can briefly describe the Petrov classification as following from
the fact that for any non-zero Weyl tensor $C_{abcd}$, as defined by
\begin{equation}
%\label{eq:3.50}
R^{ab}{}_{cd}=C^{ab}{}_{cd}-\smfrac{1}{3}R\delta _{[c}^a\delta
_{d]}^b+2\delta _{[c}^{[a}R_{\;\;d]}^{b]},
\end{equation}
there are four null vectors $k^a$, the principal null directions,
which satisfy the equation
\begin{equation}
%\label{eq:4.15}
k_{[e}C_{a]bc[d}k_{f]}k^bk^c=0.
\end{equation}
If these four vectors are distinct, one has the general case, known as
Petrov type I. The remaining cases, where two or more coincide, are
called algebraically special. If just one pair coincide, we have type
II, if three coincide, type III, and if all four coincide, type N: the
other possible degeneracy, two coinciding pairs, is called type D. The
case of zero Weyl tensor, where the spacetime is conformally flat, is
sometimes called type O.

These two developments led to the two main themes of the organization
of the exact solutions book \cite{SteKraMac03}, i.e.\ classification
by symmetry groups and algebraically special metrics.

Work on the latter area was much assisted by the development of the
already-mentioned Newman-Penrose formalism \cite{NewPen62}, a
calculation technique not based on coordinates but on a tetrad of null
vectors, and therefore well adapted to the study of algebraically
special spacetimes. These ideas led to the discovery of a number of
important solutions, for example the following.
\begin{itemize}
\item The Taub-NUT solution (or rather its NUT part):
  \cite{NewTamUnt63}.
\item The Kerr solution, the rotating black hole: \cite{Ker63}.
\item The Robinson-Trautman class: \cite{RobTra62}.
\item (Later) the Kerr-Schild ansatz and solution class:
  \cite{KerSch65}.
\end{itemize}

The excellent review of Ehlers and Kundt \cite{EhlKun62}, perhaps the
first modern general review of exact solutions, summarized the first
decade of the resulting developments.  I will now pause in describing
methods for finding solutions and discuss the uses of three of these
solutions or solution classes.

\subsection{The Taub-NUT solution}

The metric in this case can be given as
\begin{equation}
ds^{2}=\: - U^{-1}\: d\tau^{2}+\: (2\ell)^{2}\: U\: (d\psi\: + {\rm cos}\:
\theta\: d\phi)^{2}\:\:+\: (\tau^{2} + \ell^{2})\: (d\theta^{2}+ {\rm sin}^{2}
\theta\: d\phi^{2}),
\end{equation}
where $U$, which is positive in the Taub region and negative in the
NUT region, is given by
\begin{equation}
U(\tau)= -1+2\: \frac {m\tau+\ell^{2}} {\tau^{2}+\ell^{2}}.
\end{equation}
This played such a role that Misner described it as a ``counterexample
to almost anything'' \cite{Mis63}.

Papers by Misner \cite{Mis63} and Misner and Taub \cite{MisTau68}
established that the Taub and NUT regions can be joined, that the NUT
region contains closed timelike lines and no sensible Cauchy surfaces,
that there are two inequivalent maximal analytic extensions of the
Taub region (or one non-Hausdorff manifold with both extensions), that
Taub-NUT space is nonsingular in the sense of a curvature singularity,
and that there are geodesics of finite affine parameter length.

To summarize, it has the following properties:
\begin{itemize}
\item the topology of group orbits changes at the horizon;
\item there are closed timelike lines in the NUT region;
\item the boundary of the Taub region has closed null geodesics;
\item there is geodesic incompleteness at finite affine parameter
  without a curvature singularity; and
\item there are inequivalent extensions, or a non-Hausdorff one.
\end{itemize}

The solutions also have applications and generalizations outside strict
general relativity, e.g.\ in string theory or Euclidean quantum
gravity.

The solution has thus had a great influence on studies of exact
solutions and cosmological models which are spatially-homogeneous, and
more generally on those which are hypersurface-homogeneous and
self-similar (see e.g.\ the discussion in \cite{Jan87}), on cosmology
in general, and on our understanding of global analysis and
singularities in space-times.

\subsection{$pp$-waves and plane waves}

The plane waves were first found by Brinkmann \cite{Bri25} but their
significance, showing among other things that gravitational waves were
definitely not a coordinate effect, was not appreciated until the
1950s, in work of Bondi, Pirani and Robinson \cite{BonPirRob59}, Peres
\cite{Per59}, Hely \cite{Hel59} and others. These are the metrics
\begin{equation}
% \label{eq:21.38}
\df s^2=2\df\zeta \df\bar \zeta -2\df u \df v-2H\df u^2,
\quad H=H(\zeta ,\bar \zeta ,u),
\end{equation}
with $H=A(u)\zeta ^2+\overline{A}(u)\bar \zeta ^2+B(u)\zeta \bar
\zeta$. They are members of the more general $pp$-wave class, which
are given, for electrovacuum cases, by
\begin{equation}
%\label{eq:21.42}
H=f(\zeta ,u)+\bar f(\bar \zeta ,u)+\kappa _0F(\zeta ,u)%
\overline{F}(\bar \zeta ,u),\quad F_{ab}=2k_{[a}F_{,b]},
\end{equation}
where the functions $f$ and $F$ are arbitrary functions analytic in
$\zeta $ and dependent on the retarded time coordinate $u$. Here
$\kappa _0$ is the constant in the Einstein equations.

Among their properties are:
\begin{itemize}
\item the wave front speed is the speed of light;
\item they have a transverse character;
\item there are focussing effects leading to caustics;
\item there is no global Cauchy surface (i.e.\ no initial value
  problem).
\end{itemize}

These solutions are comprehensively discussed in the review of Ehlers
and Kundt \cite{EhlKun62}, so the treatment in the exact solutions
book \cite{SteKraMac03} adds virtually nothing. They have interesting
singularity and horizon behaviour, and gave rise to a number of
special cases of interest, such as sandwich waves and impulsive
waves. One can also study collisions between them
\cite{Pen65,Sze70,Sze72,KhaPen71} and the resulting metrics have been
studied in the monograph of Griffiths \cite{Gri91}: see also Ch.\ 25
of \cite{SteKraMac03}.

They have more recently been used in understanding higher-dimensional
theories and as examples in quantum field theory (they have the
property of having no quantum corrections).

\subsection{The Kerr solution}

Together with the Robertson-Walker and Schwarzschild solutions, this
is probably the best-known exact solution, because it represents the
unique rotating vacuum black hole. It has a number of interesting
mathematical properties, having indeed been found as the complement to
the NUT investigation of Petrov type D vacuum solutions (the NUT
solutions had initially been thought to be the only such solutions).
The metric is
\begin{eqnarray}
\df s^2 & = & \left( 1-
\dfrac{2mr}{r^2+a^2\cos ^2\vartheta }\right) ^{-1}  \bigg[
(r^2-2mr+a^2)\sin
^2\vartheta \df \varphi ^2 \nonumber \\
&  &
+\left( r^2-2mr+a^2\cos ^2\vartheta \right) \bigg(
\df \vartheta ^2+\dfrac{\df r^2}{r^2-2mr+a^2}\bigg) \bigg]
%\label{eq:18.25}
 \\
&  & -\left( 1-\dfrac{2mr}{r^2+a^2\cos ^2\vartheta
}\right) \bigg( \df t+\dfrac{2mar\sin ^2\vartheta \df \varphi }
{r^2-2mr+a^2\cos ^2\vartheta }\bigg)^2 . \nonumber
\end{eqnarray}

When the global structure of this metric was worked out, it emerged
that it has a ring singularity (for $m>a$) through which further
exterior regions can be connected: see e.g.\ the summary in
\cite{Car72}. Moreover, the Hamilton-Jacobi and Klein-Gordon equations
are separable in this metric, which is related to the fact that it has
a non-trivial Killing tensor. It exhibits the phenomenon of an
ergosphere, a region outside the black hole horizon but within which
any particle has to corotate around the hole. There is a relation to
work on characterization of stationary axisymmetric spacetimes by
multipole moments: the Kerr solution has very specific relations
between its moments which do not appear to be found in physical bodies
of rotating fluid, posing a question about possible sources or the
process of approach to the Kerr solution as the eventual black hole
outcome of a collapse.

Work on its mathematical properties, however, is likely to be exceeded
by the many papers on its astrophysical implications. For example, the
ergosphere classically allows the Penrose process, in which a part of
a body dividing within the ergosphere can emerge with more energy than
the original body entered with. The wave version of this is
superradiance, where the scattered wave has more energy than the
incident wave, and this phenomenon was one of the stimuli to the laws
of black hole mechanics later explained by Hawking. It is also related
to the Blandford-Znajek mechanism \cite{BlaZna77}, in which a magnetic
field threading the black hole can extract rotational energy from the
hole.

The most important of all the astrophysical uses of the Kerr solution
is probably as a basis for accretion disk physics, thought, for
example, to be responsible for the X-ray emission of X-ray binaries in
the sky, and used in explanations of larger objects such as jets in
active galactic nuclei.  Observations of astronomical accretion disks
are now suggesting the objects at their centres really are Kerr black
holes, as the only way to explain both their short periods and other
orbital data \cite{GenSchOtt03,MilFabRey04}.  Chandrasekhar
remarked\footnote{This remark came to my attention in Bicak's review
  \cite{Bic00}.} ``In my entire scientific life, extending over
forty-five years, the most shattering experience has been the
realization that an exact solution of Einstein's equations, discovered
by the New Zealand mathematician Roy Kerr, provides an absolutely
exact representation of untold numbers of massive black holes that
populate the Universe...''

\subsection{Finding more solutions: generating techniques}

A third set of novel techniques appearing for the first time in papers
in the 50s and early 60s (by Buchdahl, Ehlers, and Bonnor, for
instance \cite{Buc54,Ehl57,Bon61}) took rather longer to grow to
maturity than the first two methods described in this section: this
third topic is that of the generating techniques. Although they exist
for metrics with one Killing vector, they are most used for stationary
axisymmetric solutions, and the other classes with two commuting
Killing vectors: cylindrical waves and colliding plane waves,
boost-rotation symmetric spacetimes and cosmologies with two commuting
spacelike Killing vectors. They work not only for vacuum, but for
other forms of matter with characteristic propagation speed equal to
the speed of light: massless scalar fields (or 'stiff fluid' in the
case of a timelike gradient of the field), (massless) neutrinos, and
electromagnetism.

This area exploded in the 1970s and 1980s in work of many people. The
number of methods proliferated, and created a secondary industry of
understanding the relations between them. At the same time the number
of applications grew hugely. There is so much work that I shall not
attempt to give references here but instead refer the reader to
\cite{SteKraMac03} (see also \cite{Gri91} and \cite{BelVer01}). The best known
formulation for the basic equations is that due to Ernst. For
the stationary axisymmetric metrics
\begin{equation}
% \label{eq:17.15}
\df s^2=\e^{-2U}(\gamma _{MN\vsa}\df x^M\df %
x^N+W^2\df \varphi ^2)-\e^{2U}(\df t+A\df \varphi )^2,
\end{equation}
 where
the metric functions $U$, $\gamma _{MN}\,$, $W$, $A$ depend only on
the coordinates $x^M=(x^1,x^2)$ which label the points on the
2-surfaces $S_2$ orthogonal to the orbits, and the electromagnetic
field is given by a complex potential $\Phi$,  it reads
\begin{eqnarray}
({\rm Re}\,{\cal E}+\Phi\overline{\Phi})W^{-1}(W{\cal E}_{,M})^{;M}
 &=&{\cal E}_{,M} ({\cal E}^{,M}+2\overline{\Phi}\Phi^{,M}), 
% \label{eq:17.26}
 \\
({\rm Re}\,{\cal E}+\Phi\overline{\Phi})W^{-1}(W\Phi_{,M})^{;M}&=&\Phi_{,M}
({\cal E}^{,M}+2\overline{\Phi}\Phi^{,M}). %\label{eq:17.27}
\end{eqnarray}
To recover the metric one needs the equations
\begin{eqnarray}
A_{,M} & = & W\e^{-4U}\varepsilon_{MN\vsa}\omega^N ,\quad
\omega_N={\rm Im}\,{\cal E}_{,N}-\ii(\overline{\Phi}\Phi_{,N}-
\Phi\overline{\Phi}_{,N}), %\label{eq:17.30}
 \\
\e^{2U}& =& {\rm Re}\,{\cal E}+\overline{\Phi}\Phi,%\label{eq:17.31}
\end{eqnarray}
where $\varepsilon _{MN\vsa}$ is the Levi-Civita tensor in the $S_2$.

Although I will not attempt a review here of the methods, the works of
Geroch, Neugebauer and Kramer, Hoenselaers, Kinnersley and
Xanthopoulos (HKX), Kinnersley and Chitre, Harrison, Belinski and
Zakharov, Hauser and Ernst, Yamazaki, and Cosgrove were so influential
their names should be mentioned.

These methods themselves can now be seen as embedded in an even more
general context of symmetries of differential equations and integrable
systems, involving concepts such as inverse scattering and Lax pairs,
B\"acklund transformations, Riemann-Hilbert problems, prolongation and
so on (see \cite{MasWoo96} or Ch.\ 10 of \cite{SteKraMac03}). Thus the
solutions have offshoots in mathematics and in other physical
theories. They provide a unification of results on known solutions
(for example, all stationary axisymmetric electrovacuum solutions in
which a portion of the rotation axis is regular can be generated from
flat space).

They also enable an infinite number of solutions to be obtained. At
one stage there were a significant number of papers which exploited
this possibility by exhibiting specific solutions, but it quickly
became apparent that merely obtaining a new solution, when it can be
done infinitely often (at least in principle), is pointless. Attention
is now normally directed to ways to generate solutions with
predetermined characteristics: for example one can ask what class of
axis data gives a certain feature to the solution?

Of the solutions obtained or obtainable by these methods, a number
have been of importance or interest, for example the Tomimatsu-Sato
family, the double Kerr solution, the Neugebauer-Meinel dust disk, and
a number of colliding wave and cosmological solutions. Brevity
precludes detailed discussion and the reader is referred to the
literature already cited.

\section{Finding and using more solutions: recent developments}

I now return to the fundamental question raised earlier: how can we
compare solutions? In particular how can we test for local isometric
equivalence? This is called the equivalence problem, and belongs to
the general class of recognition problems. It has largely been
answered, in theory and in practice, using ideas due to Cartan, Brans,
Karlhede and others, with practical implementation and development by
{\AA}man, and later myself and my group, although in a formal sense its final
step is undecideable. I will briefly describe the procedure now: a full
description would be a lecture in itself. For a review see Ch.\ 9 of
\cite{SteKraMac03}.

The key point is that scalar polynomial invariants, i.e.\ polynomials
in the Riemann tensor and its derivatives in which all indices are
contracted over, are insufficient. Instead we need curvature
invariants of a more general sort, the `Cartan invariants'. These are,
for example, given by the components of the Weyl tensor referred to a
tetrad chosen using the principal null directions.

These enable local characterization of the metric. The basic idea is
to use invariantly defined tetrads, like the one from the principal
null directions, and take components of the Riemann tensor and its
derivatives in this frame. Thus the method has links with the Petrov
classification and tetrad methods which were introduced in the 1950s
and 1960s.  Counting functionally independent invariants gives the
dimension of the symmetry group, thus linking to the group theoretic
ideas of that period, and additional information available can give
the group structure.

These characterization methods can be used to check if solutions found
are really new, or to search among known solutions for examples with
desired local properties. They could in principle be used to find
solutions, as well as classify known ones, and first examples of this
method have been developed by Bradley, Karlhede and Marklund.

Other uses of this approach, i.e. the direct use of invariants, have
been in understanding the limits of families of solutions without
needing trial and error for the appropriate coordinate transformations
(for example, studying the limits of the Schwarzschild family as $m
\rightarrow \infty$); proving (non) existence of matchings by
characterizing the geometries of the proposed matching surfaces (e.g.\ 
in work of my student Daniel Cox \cite{Cox03}); and in providing a
method to `unravel' directional singularities (in work of my student
John Taylor \cite{Tay05}). The classifying quantities might also be
used to give a topology on the space of solutions, which could even be
of interest in numerical relativity.

Summarizing, the main methods for finding solutions in current use are
still those outlined in Section 4, but the understanding and
classification of these solutions can now be done in an invariant
manner which should enable better use of the solutions found and may
also provide a fresh and more invariant way to find more.

\section*{Appendix}

The second edition of the exact solutions book contains about 400
pages of new material, covering hundreds of new solutions and
references. In its preparation the authors read about 4000 new papers
(as well as the 3000 read for the first edition). So there is far more
material than I could talk about. Most of the new material is
integrated into and expands existing chapters and sections: additional
sections were added on the GHP formalism and other calculi, junction
conditions and so on, and the chapter on solutions obtained by
generating techniques was almost completely re-written.

The entirely new chapters or part-chapters were on:
\begin{itemize}
\item Homotheties
\item Characterization by invariants
\item Generating techniques themselves
\item Dynamical systems methods
\item Inhomogeneous solutions with two spacelike Killing vectors
\item Colliding plane waves
\item Special vector and tensor fields.
\end{itemize}

%%%%%%%%%%%%%%%%%%%%%%%%%%%%%%%%%%%%%%%%%%%%%%%%
%% BACKMATTER
%%%%%%%%%%%%%%%%%%%%%%%%%%%%%%%%%%%%%%%%%%%%%%%%

%\begin{theacknowledgments}
%  Infandum, regina, iubes renovare dolorem, Troianas ut opes et
%\end{theacknowledgments}




%\bibliography{ere05}


\end{document}